\documentclass[%
preprintnumbers,
nofootinbib,
 amsmath,amssymb,
 aps,
]{revtex4-1}

\usepackage{xcolor}
\usepackage{graphicx}
\usepackage{dcolumn}
\usepackage{bm}
\usepackage{hyperref}

\usepackage[normalem]{ulem} 

\newcommand{\kp}{\kappa}
\newcommand{\lm}{\lambda}

\newcommand{\vep}{\varepsilon}

\newcommand{\vp}{{\mathbf{p}}}

\renewcommand{\S}{{S}}

\newcommand{\la}{\langle}
\newcommand{\ra}{\rangle}

\begin{document}

\preprint{KCL-2020-54}

\title{Scalar resonance in graviton-graviton scattering at high-energies: the graviball  }

\author{D. Blas}
\email{diego.blas@kcl.ac.uk}
\affiliation{Theoretical Particle Physics and Cosmology Group, Department of Physics,\\
 King's College London, Strand, London WC2R 2LS, UK.}

\author{J.~Martin Camalich}
\email{jcamalich@iac.es}
\affiliation{Instituto de Astrof\'{\i}sica de Canarias, C/ V\'{\i}a L\'actea, s/n E38205 - La Laguna, Tenerife, Spain.\\
Universidad de La Laguna, Departamento de Astrof\'{\i}sica, La Laguna, Tenerife, Spain.}

\author{J.~A. Oller}
\email{oller@um.es}  
\affiliation{Departamento de F\'{\i}sica. Universidad de Murcia. E-30071,
 Murcia, Spain.}

\date{\today}

\begin{abstract}
We study graviton-graviton scattering in partial-wave amplitudes after unitarizing their Born terms.
In order to apply $S$-matrix techniques, based on unitarity and analyticity, we introduce an $S$-matrix associated  to this resummation that is free of infrared divergences.
This is achieved  by removing the diverging phase factor calculated by Weinberg that multiplies the $S$ matrix, and that stems  from the virtual infrared gravitons. A scalar graviton-graviton resonance with vacuum quantum numbers ($J^{PC}=0^{++}$) is obtained as a pole in the nonperturbative $S$-wave amplitude, 
which we call the {\it graviball}.   Its resonant effects along the physical real-$s$ axis may
peak at values substantially lower than the UV cutoff squared of the theory. 
For some scenarios, this phenomenon could have phenomenological consequences at relatively low-energy scales, similarly to the $\sigma$ resonance in QCD.
\end{abstract}

\maketitle

\section{Introduction}

Quantizing gravity remains one of the main
open problems in fundamental physics since the development of general relativity and quantum mechanics more than a century ago. The framework of quantum field theory, on the other hand, allows one to gain valuable insights starting from scattering amplitudes with given initial and final states.
Indeed, its application to scattering in gravitational theories has a long history of remarkable results, with recent vigorous progress~\cite{Weinberg:1964ev,Weinberg:1964ew,Weinberg:1965rz,Weinberg:1965nx,Bern:2002kj,Giddings:2011xs,Porto:2016pyg,Cheung:2017pzi,Strominger:2017zoo,Donoghue:2017pgk,Ciafaloni:2018uwe}.
The study of scattering processes is usually framed in terms of the $S$-matrix,  a concept first introduced by Wheeler within the context of a theoretical description of the scattering of light nuclei  \cite{Wheeler:1937a,Wheeler:1937b}, and settled as a fundamental entity of study by Heisenberg \cite{Heisenberg:1943I,Heisenberg:1943II}. 
A  rich direction has been the study of an $S$-matrix formulation of gravity in different regimes, where its character as a unitary operator and its further interplay with  string theory and  black-hole physics has been particularly fruitful~\cite{Gross:1968in,Amati:1987wq,tHooft:1996rdg,Amati:2007ak,Giddings:2009gj,Bezrukov:2015ufa,Dvali:2014ila,Alonso:2019ptb,Guerrieri:2021ivu}. 

The main purpose of our work is to explore a new complementary aspect in this program based on the
connection between gravity, quantum field theory and $S$-matrix theory that is obtained by exploiting a novel analysis in partial waves of gravitational two-particle scattering amplitudes. At energies sufficiently smaller than the Planck mass one can formulate gravitational amplitudes within the framework of effective field theory \cite{weinphys,Donoghue:1993eb,donoghue.200205.1,BjerrumBohr:2002kt,burgess,Donoghue:2017pgk}. As we discuss in detail in this work, the infrared divergences affecting the angular projections (needed for the calculation of the partial-wave amplitudes) of the gravitational tree-level scattering amplitudes 
at the lowest order (in the Newton constant $G$) are canceled by
an infrared divergent phase factor multiplying the  $S$ matrix (the ``Dalitz phase'' \cite{Dalitz:1951ai} calculated by Weinberg in \cite{Weinberg:1965nx}).
This does not depend on angles, being  a universal factor to all angular momenta. 
In this way, we can connect scattering in gravity with advanced nonperturbative techniques from hadron physics based on the unitarization of on-shell partial-wave amplitudes. These methods can restore exact two-body unitarity, while crossing symmetry is treated perturbatively from the point of view of the low-energy effective field theory, making it susceptible to improvement order by order. See~\cite{orev} for a recent review, and Ref.~\cite{Salas-Bernardez:2020hua} for an estimate of the systematic effects associated with this perturbative treatment of crossing symmetry in Chiral Perturbation Theory and in the Higgs Effective Field Theory.

A remarkable feature of the strong interactions is the generation of resonances once the lowest-order tree-level amplitudes in the associated low-energy effective field theory are unitarized,  and as result of numerical enhancements affecting some attractive tree-level partial-wave amplitudes \cite{nd}. 
An outstanding example is the $\sigma$, which is the lightest resonance in quantum chromodynamics (QCD) \cite{pdg,Oller:1997ti,Dobado:1996ps,Caprini:2005zr,Pelaez:2015qba} and whose pole position $s_\sigma$ in the complex-$s$ plane  is significantly smaller than the cutoff squared of the theory (we employ the standard notation $s$, $t$ and $u$ for the usual Mandelstam variables). The history of this resonance, also called the $f_0(500)$, is fascinating \cite{Pelaez:2021dak}, being finally recognized by the Particle Data Group (PDG) since 2002 \cite{pdg}. This
was possible thanks to the relation of the $\sigma$ pole with chiral dynamics, as clearly established by unitarization methods of Chiral Perturbation Theory \cite{Oller:1997ti,nd,Dobado:1996ps,Oller:2004xm}, and/or the Roy equations \cite{pdg,Caprini:2005zr,GarciaMartin:2011jx}, as well as by new experiments from heavy-meson decays yielding two-pion event distributions  with a clear resonance signal associated to the $\sigma$ \cite{Aitala:2000xu,Asner:1999kj}. 

This resonant phenomenon affects prominently the scalar dynamics at low-energies in QCD, like the excitation of the quark condensate (the lightest-pseudoscalar scalar form factors) \cite{Colangelo:2001df,Oller:2007xd,Oller:2008kf},   the large corrections to the current algebra prediction of the isoscalar scalar-scattering lengths, and in the $S$-wave $\pi\pi$ phase shifts in general,  its above-mentioned role in the low-energy part of several two-pion event distributions from heavy-meson decays, $\pi$-nucleon $\sigma$ term \cite{Gasser:1988jt,Alarcon:2011zs}, etc~\cite{Pelaez:2015qba}. 

Given that quantum gravity and QCD admit derivative low-energy expansions in terms of appropriate effective-field theories, one is naturally led to consider the emergence of resonances from nonperturbative gravitational interactions with similar unitarization tools as in QCD. 
The main result of this work is that we indeed find a new resonance pole in the graviton-graviton $S$-wave partial-wave amplitude, with vacuum quantum numbers ($J^{PC}=0^{++}$). There is a remarkable analogy between this resonance in gravity and the $\sigma$ in the chiral limit of QCD, in the sense that the position of the poles is  similar relative to the cutoffs of their respective effective field theories.
We call this resonance the {\it graviball}~\footnote{The term graviball was recently used in~\cite{Guiot:2020pku}. This semiclassical analysis  has no relation with our pure  quantum-mechanical resonance in two-graviton scattering. In self-completeness scenarios of gravity \cite{Dvali:2010bf,Dvali:2011aa} unitarity also requires the production of black holes composed by gravitons that may be interpreted as resonances, and the graviball could be the lightest one among them \cite{Dvalines}.} given the resemblance  with the \textit{glueballs} of the pure Yang-Mills theory~\cite{Fritzsch:1975tx,Bali:1993fb,Morningstar:1999rf,Lucini:2004my}. Let us also note that our unitarization method is in partial wave amplitudes and applies at energies  $|s|<G^{-1}$, with typically $s\sim t$, while the much studied eikonal approximation provides unitary amplitudes in a very different scenario for $|s|\gg G^{-1}$ at small angles $|t/s|\ll 1$ \cite{Amati:2007ak,Kabat:1992tb}.

Our results also imply that in theories with a  large number of light degrees of freedom~\cite{Dvali:2007hz,Dvali:2007wp,Arkani-Hamed:2016rle}, which may naturally appear in models of large extra dimensions \cite{ArkaniHamed:1998rs,Antoniadis:1998ig,Csaki:2004ay},  
the position of the pole becomes 
$\ll G^{-1}$ and its effects may be relevant for scales probed during inflation (or even at much lower ones  \cite{Dvali:2008fd}).  This may open up several phenomenological consequences 
like in studies of primordial gravitational waves or inflation.    
Furthermore, in these models, the resonance may also modify the gravitational phenomena at scales as low as $\mu m$ \cite{ArkaniHamed:1998rs,Dvali:2001gx}  (corresponding to a UV cutoff around the TeV scale).

As a side result, it is also worth stressing that the infrared divergences, associated with massless force carriers like gravitons or photons, affect other processes too (a prominent example is the interference between the strong and  Coulomb interactions). Therefore, our approach to tackle infrared divergences  of QED in partial waves is also called to play an important role in studies based on the unitarization of strong interactions involving charged particles~\cite{Alarcon:2012kn,Guo:2012vv}.

\section{Framework}

To address the questions above, we deal with the quantum formulation of  gravity
within the foundations of effective field theory (EFT) where gravitational interactions are organized in a derivative  
expansion \cite{weinphys,Donoghue:1993eb,donoghue.200205.1,BjerrumBohr:2002kt,burgess,Donoghue:2017pgk}. 
Weinberg in his seminal paper \cite{Weinberg:1965nx} showed that the whole effect of virtual infrared gravitons (as well as for photons in QED) is a multiplicative factor affecting the $S$ matrix (see also  \cite{Giddings:2009gj}) whose phase is irrelevant for physical observables and it is independent of spin, and of angle in two-body scattering,  cf. its Sec.~V (see also~\cite{Dalitz:1951,Kulish.200121,Ware:2013zja}). As we show below, by eliminating this phase one can properly define the partial-wave amplitude expansion of the graviton-graviton scattering Born terms, and then build a new $S$-matrix in partial waves by making use of nonperturbative unitarization methods.

We focus on   the graviton-graviton scattering process $|\vp_1,\lm_1\ra |\vp_2,\lm_2\ra\to |\vp'_1,\lm'_1\ra|\vp'_2,\lm'_2\ra$, 
where we denote by $|\vp, \lm\rangle$ the one-graviton state of three-momentum $\vp$, energy $p_i^0=|\vp_i|$ and helicity $\lm$. The $S$- and $T$-matrix operators are related by 
$S=I+i (2\pi)^4 \delta^{(4)}(P_f-P_i) T$, with $P_f$ and $P_i$ the total final and initial four-momenta, respectively, and we follow the conventions of  \cite{graviballlong}.
The basic input blocks for our study   are  tree-level or Born amplitudes, that we 
 adapt from~\cite{grisaru.170513.1}.
They are given according to the number of gravitons with  helicity $-2$ and can be obtained by crossing from
\begin{equation}
\begin{split}
\label{200119.6}
F_{22,22}(s,t,u)&=\frac{\kp^2}{4}\frac{s^4}{stu}~,\\
\end{split}
\end{equation}
where $\kp^2=32\pi G$, 
and $F_{\lm'_1\lm'_2,\lm_1\lm_2}(s,t,u)$ refers to the Born scattering amplitudes.

Graviton-graviton scattering can also be formulated in the 
basis with well defined angular momentum $|pJ,\lm_1\lm_2\ra_\S$, leading to the partial-wave amplitudes (PWAs),
\begin{align}
\label{200119.4}
\bar{T}^{(J)}_{\lm'_1\lm'_2,\lm_1\lm_2}(s)\equiv {_\S\la} pJ,\lm_1'\lm_2'|T|pJ,\lm_1\lm_2\ra_\S,
\end{align}
where $J$ is the total angular momentum with third component $\mu$ and the subscript $\S$ denotes that the states have been symmetrized according to the Bose symmetry of the states \cite{graviballlong}. Since $\mathbf{J}$ and $H$ commute one can always consider such basis for the continuum spectrum of the theory. Even if the PWA series does not converge in  graviton-graviton scattering ~\cite{Lehmann:1958ita,Giddings:2009gj,martin.200705.1} (as it already happens in non-relativistic Coulomb scattering \cite{Kang:1962}), certain processes will be 
dominated by a particular PWA,  for instance, the physical effects of the {\it rescattering}
of two gravitons produced with the quantum numbers of a given partial wave (like the case of two-pion event distributions mentioned above in connection with the $\sigma$), or due to interferences with other forces (analogously to the interplay between strong and Coulomb interactions, already pointed out). 
The $S$-matrix in partial waves $\bar{S}^{(J)}(s)$ is given by~\cite{graviballlong}
\begin{align}
\label{200121.11}  
\bar{S}^{(J)}(s)&=I+i\frac{\pi 2^{|\lm|/4}}{4}\bar{T}^{(J)}(s)~,
\end{align}
with $|\lm|=|\lm'_2-\lm'_1|=|\lm_2-\lm_1|=0$ or 4. 
Unitarity, when saturated by the two-graviton cut, implies $\bar{S}^{(J)}\bar{S}^{(J)*}=1$. Contributions of the cuts given by more intermediate gravitons are suppressed by ${\cal O}((Gs)^3)$ in the low-energy EFT of gravity, and we  ignore them.  
We only consider the partial waves $T^{(J)}_{22,22}(s)$ (for even $J$) and $T^{(J)}_{2-2,2-2}(s) $, since all the other PWAs are either zero, or related to the previous ones by symmetry properties of the PWAs under the exchange of the helicity labels \cite{graviballlong}.

\subsection{Infrared singularities} 

The direct application of the previous PWA formalism to the amplitudes in Eq.~\eqref{200119.6} is hindered by divergences reflecting the infinite-range character of the gravitational interaction  \cite{Giddings:2009gj}.
For instance, the PWA of the Born amplitude $T_{22,22}(s,t,u)$ with  $J=0$ is given by 
\begin{align}
\label{200121.2}
F^{(0)}_{22,22}(s)& 
=-\frac{\kappa^2 s^2}{16\pi^2}\int_{-1}^{+1} \frac{d\!\cos\theta}{t}~,
\end{align}
which has a logarithmic divergence  for $\cos\theta\to 1$.
This is actually an infrared (IR) divergence  because the four-momentum squared $t$ 
of the $t$-channel exchanged graviton vanishes for $\cos\theta\to 1$.
It is associated to a virtual soft graviton that connects
two external on-shell graviton lines, following the classification of~\cite{Weinberg:1965nx}. This reference studies in detail this type of divergences and shows that 
they can be resummed correcting the $S$-matrix by a global phase. For our case of concern regarding the amplitudes in Eq.~\eqref{200119.6}, where the external particles are massless, this
factor reads\footnote{Take the limit of vanishing $m_n$, $m_m$ in Eq.~(5.11) of \cite{Weinberg:1965nx} and sum over the final and initial  graviton  pairs in the $s$ channel.} 
\begin{align}
\label{200914.1}
S_c(s)
=\exp\left[-i2 G s \log\frac{\mu}{\mathfrak{L}}\right],
\end{align}
where  $\mu\to 0^+$ is an IR cutoff and the three-momentum scale $\mathfrak{L}$ 
 separates the regions of ``hard'' and ``soft'' graviton momenta (chosen low enough to justify the approximation of soft momenta in a given process). 
 It is important to emphasize that $S_c(s)$  is independent of $J$, being common to all PWAs with the given set of helicities $\lm_i$ and $\lm_i'$~\cite{graviballlong}.~\footnote{This connection between the soft infrared divergences of the one-loop scattering amplitudes and the kinematic divergences in the tree-level PWAs has been independently demonstrated for gauge theories and gravity in Ref.~\cite{Baratella:2020dvw}}.

We next redefine the $S$ matrix in PWAs by extracting the divergent phase, $\bar S^{(J)}= S_cS^{(J)}$, 
which also fulfills that $S^{(J)}S^{(J)*}=1$. 
We denote by $V^{(J)}$ the partial-wave projected Born amplitudes corresponding to $S^{(J)}$, such that, analogously to Eq.~\eqref{200121.11},
\begin{align}
S^{(J)}(s)=I+i\frac{\pi 2^{|\lambda|/4}}{4}V^{(J)}+{\cal O}(G^2),
\end{align}
and from here,
\begin{align}
\label{200121.18b}
V^{(J)}(s)&=F^{(J)}(s)+\frac{1}{2^{\lm/4}}\frac{8Gs}{\pi}\log\frac{\mu}{\mathfrak{L}}~.
\end{align} 
Notice that $V^{(J)}(s)$  is free of the IR divergences~\cite{graviballlong} affecting the direct partial-wave projection of the Born term $F^{(J)}(s)$ and, instead, carries a dependence on the IR scale $\mathfrak{L}$. A change of this cutoff produces just a rescaling of the $S^{(J)}$-matrix, $S^{(J)}_{\mathfrak{L}^\prime}=S^{(J)} (\mathfrak{L}/\mathfrak{L}^\prime)^{2i G s}$. Therefore, although the dependence on $\mathfrak{L}$ is physically spurious, it shows up in a truncated perturbative expansion of $S^{(J)}$.
On dimensional grounds it is clear that 
$\mathfrak{L}\propto \sqrt{s}$, because at ${\cal O}(G)$ this is the only magnitude with dimension of momentum that is available to
the Feynman diagrams giving rise to the Born term~\cite{Weinberg:1965nx}. Therefore, we define
\begin{align}
\label{200121.20}
\mathfrak{L}=\sqrt{s}/a~,
\end{align}
where  $a$ is expected to be larger than 1, so that $\mathfrak{L}$ is appreciably smaller than $\sqrt{s}$ that fixes the external energy.
In particular, for the central $J=0$ PWA in our study one has 
\begin{align}
\label{eq:V0_dimreg}
V_{22,22}^{(0)}(s)&=\frac{8 Gs}{\pi}\log a~. 
\end{align}

\subsection{Unitarized PWAs} 

From the unitarity of $S^{(J)}$, the new $T$ matrix in partial-wave amplitudes \cite{oller.book} satisfies the  two-body unitarity relation  
\begin{align}
\label{200121.17}
\Im \frac{1}{T^{(J)}}=-\frac{\pi 2^{|\lm|/4}}{8}\theta(s)~.
\end{align}
As we saw in the previous section, {the partial-wave projected Born terms are free from any cut, cf. Eq.~\eqref{eq:V0_dimreg},} and the corresponding $T$ matrix   only has the unitarity or right-hand cut (RHC) {as a result of} Eq.~\eqref{200121.17}.  
This enables a simple parameterization of the PWAs which fulfills exact two-body unitarity   \cite{nd,Oller:2000fj},   
\begin{align}
\label{200530.1}
T_{\lm'_1\lm'_2,\lm_1\lm_2}^{(J)}(s)&=
\left[R_{\lm'_1\lm'_2,\lm_1\lm_2}^{(J)}(s)^{-1}+
g(s)\right]^{-1}~.
\end{align} 

In the previous equation  $g(s)$ is an analytical function in the cut complex-$s$ plane with only  RHC, and whose discontinuity is the same as that of $1/T^{J}(s)$.
To simplify the notation let us focus on the scattering with $\lm=0$. 
One can express $g(s)$ in terms of a once-subtracted dispersion relation, with the result~\cite{Oller:2000wa,oww,graviballlong} 
\begin{align}
\label{200531.1}
g(s)=a(s_0)+\frac{1}{8}\log\frac{-s}{s_0}~.
\end{align}

The other ingredient in the right-hand side of Eq.~\eqref{200530.1} is $R_{\lm'_1\lm'_2,\lm_1\lm_2}^{(J)}(s)$, that  has no two-body unitarity cut because this is fully accounted for by $g(s)$.
In the following we suppress the helicity subscripts of $T^{(J)}$ and $R^{(J)}$. 
Following  Refs.~\cite{oww,Oller:2000fj} we obtain $R^{(J)}(s)$ by matching  $T^{(J)}(s)$ in Eq.~\eqref{200530.1} to its perturbative calculation order by order within the low-energy EFT for gravity. Then, at LO one has 
\begin{equation}
\label{200530.2}
R^{(J)}=V^{(J)}+{\cal O}\left((G s)^2\right).
\end{equation}
Later we will also discuss the impact in our results of higher-order tree-level contributions from monomials in pure gravity involving six and eight derivatives. These are the so-called $R^3$ and $R^4$ mononomials that stem from the product of three and four Riemann curvature tensors, respectively, and are expected to appear as counterterms in   UV completions of general relativity \cite{vanNieuwenhuizen:1976vb,AccettulliHuber:2020oou}.

The issue of determining {\it a priori} a value for the subtraction constant $a(s_0)$ at the scale $s_0$ has been already discussed in QCD, e.g. regarding the $\sigma$ or $\Lambda(1405)$ resonances \cite{Oller:2000fj,Hyodo:2008xr}. 
The point is to compare  $g(s)$ with its result evaluated
with an UV cutoff $\Lambda^2$,
$g_c(s)=\frac{1}{8}\log\frac{-s}{\Lambda^2}+\mathcal O(s/\Lambda^2)$. 
Identifying $s_0$ with the cutoff squared 
and matching the expressions for $g(s)$ and $g_c(s)$, it follows that $a(\Lambda^2)=0$~\cite{graviballlong}. More intuitively, the idea behind this is to fix $a(\Lambda^2)$ such that only distances longer than $\Lambda^{-1}$ are effectively described by the propagation of the intermediate states, while preserving the right analytical properties of $g(s)$. 

Combining Eqs.~\eqref{200530.1} and \eqref{200530.2} the unitarized amplitude can be expanded to next-to-leading order as,
\begin{align}
\label{eq:TJNLO}
T^{(J)}=V^{(J)}\left(1-\frac{V^{(J)}}{8}\log\frac{-s}{\Lambda^2}\right)+\mathcal O\left((Gs)^3\right).    
\end{align}
A value of the cutoff $\Lambda$ can be thus estimated by assuming that the correction has the expected size in the EFT expansion,  $s/\Lambda^2$.
 For definiteness let us take $V^{(J=0)}(s)$, Eq.~\eqref{eq:V0_dimreg}, and then we find that the \textit{unitarity} cutoff is ${\Lambda}^2_U=\pi (\log a \,G)^{-1}$.

It is important to point out that cutoffs derived from unitarity considerations within the EFT do not correspond, necessarily, to the \textit{fundamental} scale of its ultraviolet completion \cite{donoghue.200204.1}. 

\section{The graviball}

Poles of PWAs correspond to bound states, anti-bound (also called virtual) states and resonances \cite{Badalyan:1982}. The first ones lie in the physical or first Riemann sheet (RS) and we do not find them in our present study.  Both resonances and anti-bound states occur as poles in the second or unphysical RS. To look for them, the expression \eqref{200530.1}, can be analytically continued to the second RS, so that $g(s)$ in the 2nd RS becomes $g_{II}(s)=g(s)-i\pi/4$. 

Let us discuss the pole positions in $J=0$. 
The PWA $T^{(0)}_{II}(s)$ has a resonance pole in the second RS at $s_P$, whose position satisfies
the secular equation,  
\begin{align}
\label{200206.1}
\frac{1}{\omega x}&+\log (-x)-i2\pi=0,
\end{align}
where $ x=s_P/\Lambda^2$, and
\begin{align}
\label{210309.1}
\omega=\frac{\Lambda^2}{\Lambda_U^2}=\Lambda^2 \frac{G\log a}{\pi}
\end{align}
is the ratio between the fundamental and unitarity cutoffs. For $\omega=1$,
\begin{align}
\label{eq:grav.pole.app}
s_P=(0.07-i\,0.20)\,\Lambda^2\simeq-i\frac{2}{3\pi}\Lambda^2~.
\end{align} 
We call this resonance the {\it graviball}.
Let us stress that the real part of $s_P$ is significantly smaller than its imaginary part, so that the resonance effects would peak at values of $s$  substantially lower than $ \Lambda^2$.

\begin{figure}
\begin{center}
\begin{tabular}{l}
\includegraphics[width=0.6\textwidth]{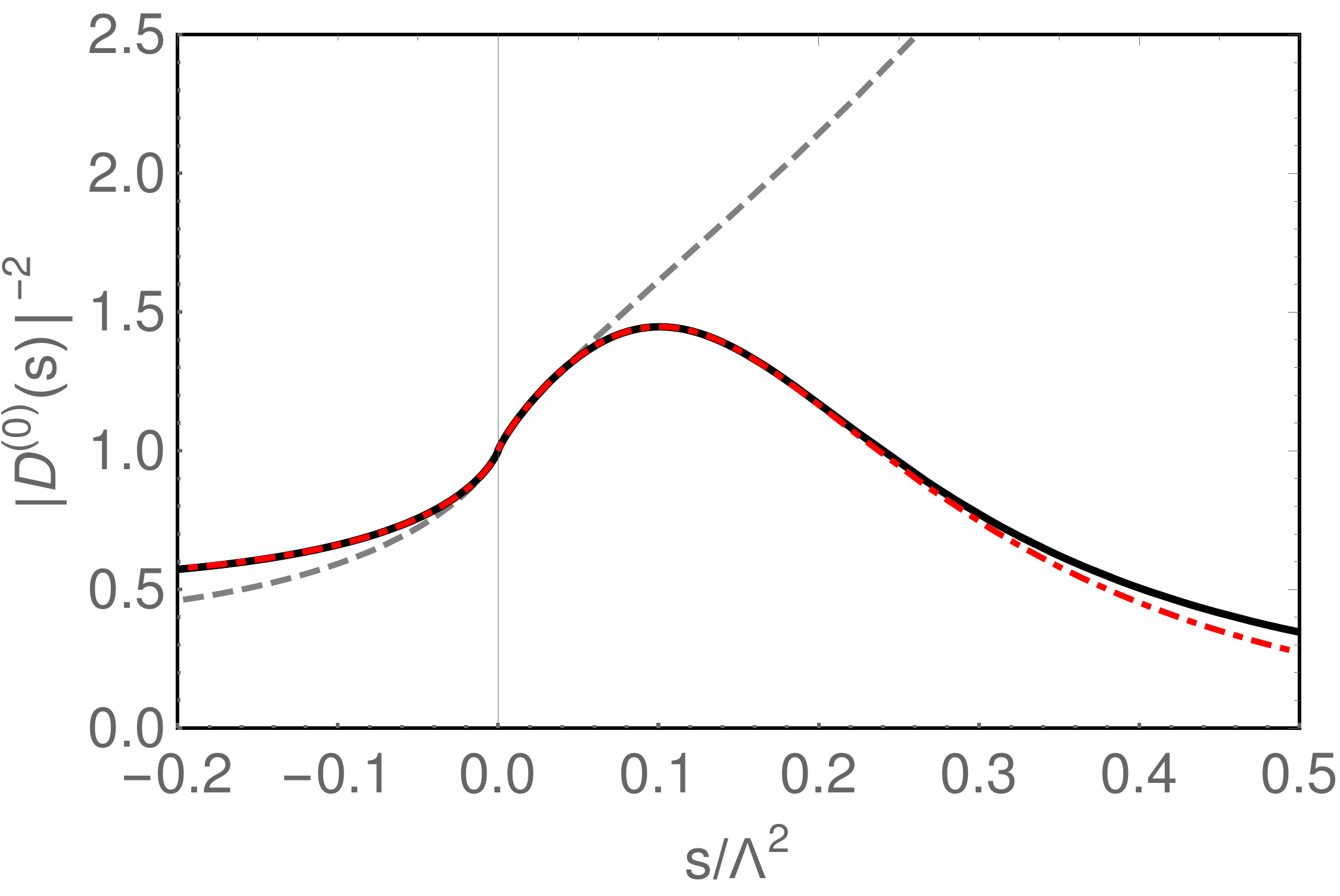} 
\end{tabular}
\caption{{\small The modulus squared of  $1/D^{(0)}(s)$  
is plotted for $J=0$  as a function of $s$ in units of $\Lambda^2$. The solid line represents the unitarized result, the dashed line is the perturbative calculation, and the dash-dotted (red) curve corresponds to the unitarized results including also an $R^4$ contribution (see main text).
\label{fig.170725.1}}} 
\end{center}
\end{figure}

This is shown in Fig.~\ref{fig.170725.1} where we plot with the solid line the modulus squared of the Omn\`es function $1/D^{(0)}(s)=T^{(0)}(s)/V^{(0)}(s)$,  which drives the final- and initial-state interactions in the corresponding graviton-graviton PWA ~\cite{oller.book}. 
For instance, one could think of a situation where multiple gravitons are produced by an energetic source so that a pair of
gravitons with quantum numbers $(J, \lm^{(\prime)}_{1-2})$ rescatters  with some energy distribution. 
This is also the function  that controls the exchange of  graviton-graviton states with such quantum numbers between initial and final states, 
or from external sources that couple to this system.  
We clearly observe a resonant shape with an $s$ dependence dominated by the  pole $s_P$ in the complex-$s$ plane,  Eq.~\eqref{eq:grav.pole.app}. It is worth stressing that a figure similar to Fig.~\ref{fig.170725.1} \cite{Oller:2004xm} enabled the direct observation of the $\sigma$ resonance in the
two-pion event distributions in the decay $D^+\to \pi^+\pi^+\pi^-$ \cite{Aitala:2000xu},  and in other heavy-meson decays, prompting the PDG to establish the $\sigma$ as the lightest resonance in QCD.

We also compare $1/|D^{(0)}(s)|^2$ in the lower-energy region of Fig.~\ref{fig.170725.1} with its perturbative expansion (dashed line) up to
linear terms in $s$, which completely fails in providing the right resonant shape  even for $s\gtrsim 0.1\Lambda^2$.  As $J$ increases, the poles move deeper into the complex-$s$ plane \cite{graviballlong} with $|s_P|/\Lambda^2$ much larger than 1, so that we only focus on  $J=0$ here. The results shown at Fig.~\ref{fig.170725.1} and the aforementioned physical effects connected to the Omn\`es function already hing towards possible phenomenological implications of the \emph{graviball}. We briefly discuss some salient ones in Sec.~\ref{sec:outlook}, and leave a more detailed analysis for future work.

An interesting application of the knowledge of the function $D(s)$ is the calculation of the probability of decay of the graviball into two gravitons as a function of time $t$. This can be expressed by the amplitude $A(t)$, that gives the probability of existence of the graviball after a time $t$ since it was created ($A(0)=1$) and is obtained by making use of basic quantum-mechanical principles of $S$-matrix theory~\cite{Goldberger:1964}:
\begin{align}
\label{220118.1}
A(t)=\int_0^\infty dE\frac{B(E^2)}{|D(E^2)|^2}e^{-i E t}~,
\end{align}
where $E$ is the total energy in the graviball rest frame and $B(s)$ is a source amplitude  associated to the production of the gravitons by the decaying state. The dependence of this function in energy is expected to be much weaker than that of $D(s)$ and it can be assumed constant and factored out from the integral.\footnote{Notice that the function $B(s)$ has no two-body RHC \cite{Goldberger:1964,Oller:2017alp}.} Comparing $|A(t)|^2$ with an exponential decay formula one can identify the width of the graviball as the point in which $|A(t)|^2=\exp(-1)$:
\begin{align}
    \label{220118.2}
    \Gamma\approx 0.23 \,\Lambda~.
\end{align} 
The $s_P$ in Eq.~\eqref{eq:grav.pole.app} is almost purely imaginary and clearly signals a situation very different to the narrow-resonance case. We then consider the application of Eq.~\eqref{220118.1} to be much more sensible than taking the square root of $s_P$, and identifying its imaginary part with $-\Gamma/2$ (as if the graviball were narrow). This naive and more crude  procedure provides a width that is more than twice as big as Eq.~\eqref{220118.2}.

\subsection{The scale $\Lambda^2$}

In the analysis of the graviball above we have assumed that the cutoff of gravity is close to the one estimated by unitarity considerations, $\omega=1$. 
Increasing $\omega$ from 1 displaces the pole closer to the real-$s$ axis and the resonance becomes lighter and narrower. For $\omega\gtrsim 10$ the real and imaginary parts of the pole are of the same order of magnitude and its absolute value monotonically decreases with $\omega$ reaching the limit $x\to0$ when $\omega\to \infty$.  
A possible way to accomplish $\omega\gg1$ is by considering  scenarios with a large number $N$ of light degrees of freedom~\cite{donoghue.200204.1,graviballlong}.
These contribute with their two-particle cuts to the unitarity relation in Eq.~\eqref{200121.17} and, thus, with a factor $N$ multiplying $g(s)$ in Eq.~\eqref{200531.1}. Ultimately, this leads to a reduction of the unitarity cutoff $\Lambda^2_U$ by a factor $1/N$. Note, though, that there are arguments suggesting that the fundamental cutoff of gravity decreases by a similar factor~\cite{han.200204.1,Veneziano:2001ah,gomez,Dvali:2007wp,Dvali:2014ila}, and one recovers $\omega\approx 1$. In any case, the graviball becomes narrower and lighter as $N$ increases, with potentially important consequences for its phenomenology. 
If the cutoff of the theory is smaller than the unitarity one then $\omega<1$, the pole moves further out in the complex-$s$ plane and for $\omega\lesssim 0.2$ its position is such that $|s_P|\gtrsim G^{-1}$, where our approach rooted in the EFT becomes clearly model dependent.

\subsection{Estimate of $\log a$}

The unitarity cutoff, and the absolute position of the pole, eventually depend on the parameter $\log a$.  
From its definition in Eq.\eqref{200121.20} one expects that $\log a\simeq\mathcal O(1)$ although it is not possible to predict its value from perturbative calculations.
One also expects that the impact on the results will diminish with higher-order unitarized PWAs~\cite{Weinberg:1965nx,Ciafaloni:2018uwe}.
Lacking them, in Ref.~\cite{graviballlong} we have developed a method to estimate an optimal $\log a$ by extending our analysis to $d>4$ inspired
by some well established methods like optimized perturbation theory \cite{Stevenson:1981vj,Brodsky:1982gc,Su:2012iy} or the dimensional $\vep$-expansion in Statistical Mechanics \cite{itzi,son,ton}. It is also important to stress that this optimal value in principle exists, and provides a value very close to the one computed non-perturbatively in a exactly solvable toy model discussed below and analyzed in detail in Ref.~\cite{graviballlong}. 

The scattering amplitude is IR-finite for $d=5$ and there should be a minimal non-integer dimension $d_c\lesssim 5$ below which the expansion in powers of $\vep=d-5$ stops being reliable.
Then, we introduce a condition of \textit{maximum smoothness} to determine $d_c$
which is based on requiring that the properties of the scattering amplitudes at $d_c$ and the regularized ones at $d=4$ are as close  as possible to each other. We implement this by defining a measure of the differences $r=|\Lambda_{d_c}^2-\Lambda^2|/\Lambda^2$, with $\Lambda_{dc}$ the cutoff at $d=d_c$ ($\omega=1$), and by minimizing it with respect to $\log a $ (for more details see \cite{graviballlong}). 
Interestingly, $r$ has a minimum for $\log a\approx 1$, 
in agreement with our expectations that $\log a={\cal O}(1)$. According to Eq.~\eqref{eq:grav.pole.app} this leads to a  pole at $d=4$ whose absolute position is
\begin{align}
\label{eq:graviball.pole}
s_P\simeq (0.22-i\,0.63)~G^{-1},     
\end{align}
providing support for the existence of a subplanckian graviball in pure gravity. In addition, for $\log a\simeq 1$ the unitarity cutoff $\Lambda\simeq \pi G^{-1}$, and a dimensional estimate of the higher-order corrections to $s_P$ gives a typical size $|x|\simeq 2/3\pi\approx 20\%$ for the one-loop corrections entering in $F_{22,22}(s,t,u)$.

In this regard, we can be more quantitative and
estimate higher-order effects by considering the contributions to graviton-graviton scattering from the monomials involving the product of three  and four Riemann tensors \footnote{The monomials of two Riemann tensors do not contribute to the $S$-matrix of pure gravity, see e.g.~\cite{AccettulliHuber:2019jqo} and refs therein.}, respectively. The former monomial
(labeled as $R^3$) involves six derivatives while the latter 
(there are three of them \cite{AccettulliHuber:2020oou}, denoted by $R^4$) have eight derivatives.  The contribution from the $R^3$ counterterm  vanishes for the $F_{22,22}$ amplitude~\cite{vanNieuwenhuizen:1976vb}. 
Hence, the only non-vanishing contribution to the graviball stems from the $R^4$ terms, whose contributions  for the graviton-graviton scattering amplitudes appeared  in  Ref.~\cite{AccettulliHuber:2020oou}, which makes use of the spinor formalism. In terms of the  Mandelstam variables  the  explicit expression for the contribution to $F_{22,22}(s,t,u)$ from this source, that we call $F_{R^4;22,22}(s,t,u)$, is
\begin{align}
\label{211228.1}
F_{R^4;22,22}(s,t,u)=\frac{\widetilde{\beta}\kappa^2}{\pi} s^4~,
\end{align}
 following the notation of Ref.~\cite{AccettulliHuber:2020oou} for the 
 counterterrm $\widetilde{\beta}$. This amplitude only gives contribution to $S$-wave and its $J=0$ partial-wave projection, $V_{R^4;22,22}^{(0)}$, is
\begin{align}
\label{211228.2}
V_{R^4;22,22}^{(0)}=\frac{8\widetilde{\beta} G}{\pi^2}s^4=\frac{8\widetilde{\beta}}{\pi \log a}\frac{s^4}{\Lambda^2},
\end{align}
where we have used $\omega=1$ 
($\Lambda=\Lambda_U$). The counterterm $\widetilde{\beta}$ is written in Ref.~\cite{AccettulliHuber:2020oou} as $\widetilde{\beta}=4(\beta_1+\beta_3)$,  with $\beta_1,\beta_3>0$ being the dimensionful  coupling constants that directly appear in the action. We  assume naive-dimensional analysis to estimate these couplings as $\beta_1,\beta_3\sim \Lambda^{-6}$. Then, up to ${\cal O}(1)$ factors, Eq.~\eqref{211228.2} is finally written as
\begin{align}
\label{211228.2}
V_{R^4;22,22}^{(0)}=\frac{32}{\pi \log a}\frac{s^4}{\Lambda^8}~.
\end{align}
The kernel $R^{(0)}$ in Eq.~\eqref{200530.2} acquires an extra term, and we explicitly have now 
\begin{align}
R^{(0)}_{22,22}(s)&=\frac{8s}{\Lambda^2}+\frac{32s^4}{\pi \Lambda^8}+\ldots
\end{align}
where we have taken $\log a=1$, and the ellipsis indicates other higher-order terms starting from one-loop contributions. In terms of $R^{(0)}_{22,22}(s)$ one can calculate the resulting unitarized $T^{(0)}_{22,22}(s)$ partial-wave amplitude  by applying Eq.~\eqref{200530.1}, and the associated $1/D^{(0)}_{22,22}(s)=T^{(0)}_{22,22}(s)/R^{(0)}_{22,22}(s)$ function. Now, the secular equation with $\omega=1$ is 
\begin{align}
(x+\frac{4x^4}{\pi})^{-1}+\log(-x)-i2\pi=0~,
\end{align}
which gives $s_P=(0.07-i\,0.21)\,\Lambda^2$. We then see that the corrections are very small, at the level of a $3\%$, which 
agrees with the expectation of having a correction in $s_P$ with a relative uncertainty of order  $|x|^3\sim 1\%$ given by an operator with eight derivatives. This is of course much smaller than the leading loop corrections that are expected to come from the one-loop calculation of graviton-graviton scattering, which relative uncertainty in $s_P$ is ${\cal O}(|x|)$, as mentioned above. These corrections, still smaller than the leading effect, will be considered in future work, together with the inclusion of matter effects.

Coming back to the choice of the factor $\log a$,  in Ref.~\cite{graviballlong} we have successfully checked  the previous method to estimate $\log a$ by applying it to a non-trivial quantum mechanical toy model, which is exactly solvable and shares key features with the analyzed case of the graviball. This model,  that we called the Adler-Coulomb scattering,  has a  Coulomb-like Born term 
times $E^2/M^2$, where  
$E$ is the non-relativistic kinetic energy
and $M$ is a UV mass scale.
The $J=0$ PWA is the same as $F^{(0)}_{22,22}(s)$ in Eq.~\eqref{200121.2}, except for a global factor. By dealing with the IR  divergences and unitarizing the model as we did before, our method predicts that $\log a\approx 1/2$, which is precisely the right value to reproduce the lightest pole positions of the exact solution.
From a phenomenological point of view, in connection with the discussion above on the scale $\Lambda^2$,  
the actual value of $\log a\simeq {\cal O}(1)$ is of limited importance to give  $s_P$ in absolute terms  compared with the (as yet unknown) number of light fields $N$ and how they affect $\Lambda$. 

 Finally, let us note that the existence of a scalar resonance
in higher space-time dimensions \cite{graviballlong} has also been identified 
within string theory~ \cite{Guerrieri:2021ivu}. In this respect we mention that  Ref.~\cite{Guerrieri:2021ivu} estimated the Wilson coefficient  $\alpha$ in ten dimensions (which controls the leading UV completion of maximal supergravity), by applying $S$-matrix bootstrap \cite{EliasMiro:2019kyf,Guerrieri:2020bto}. It  is found that the integral driving this calculation is  largely dominated by a scalar resonance, tentatively identified  with the graviball in ten dimensions in the same reference (see also the Appendix C in the supplementary material of Ref.~\cite{Guerrieri:2021ivu}).\footnote{Reference~\cite{Guerrieri:2021ivu} claims in the Appendix I of its supplementary material that  they use a  method similar to ours to unitarize perturbative amplitudes. However, we notice that  their Eq.~(I4) lacks the logarithmic branch point at threshold ($s=0$), which is kept by the function $g(s)$ within our formalism. This point is  discussed in the literature since long time ago \cite{Schnitzer:1970wa,Lehmann:1972kv}. We refer to  Ref.~\cite{Oller:2020guq} for a recent review on the unitarization techniques in hadron physics, also offering a historical account on them.} 

\subsection{Analogy with the $\sigma$}

As indicated above, there are important  similarities  
between the graviball and the $\sigma$ that we explain now.  The interactions among pions, 
as well as those for gravitons,
are  constrained by derivative low-energy EFTs
\cite{weinphys,Weinberg:1966fm,Donoghue:1993eb,donoghue.200205.1,BjerrumBohr:2002kt,burgess,Donoghue:2017pgk,Gasser:1983yg,Pich:1995bw}. 
The $S$-wave isoscalar $\pi\pi$ scattering amplitude in the chiral limit at LO is $V^{(0)}_{\pi\pi}(s)=s/f_\pi^2$, with $f_\pi$ the weak pion decay constant. 
This PWA is analogous to $V^{(0)}(s)$ in Eq.~\eqref{eq:V0_dimreg}.
Its unitarization~\cite{orev,graviballlong}, by an expression equivalent to Eq.~\eqref{200530.1},  leads to an
$s_\sigma$ that fulfills the same secular equation as  Eq.~\eqref{200206.1} in terms of $x_\sigma=s_\sigma/\Lambda^2_\pi$. For $\pi\pi$ scattering  $\Lambda_\pi=4\pi f_\pi$ is both
the unitarity and fundamental cutoff of the pion EFT \cite{Georgi:1984}.
In absolute units the pole position lies at $s_\sigma=0.08-i\,0.23~\rm GeV^2$, that is located  deep in the complex-$s$ plane (but still $|s_\sigma|/(4\pi f_\pi)^2\approx 0.17$) and has a small real part (features that persist almost untouched when taking the physical pion  masses \cite{graviballlong,Hanhart:2008mx,Albaladejo:2012te}).
This implies that its resonance effects peak for $s\ll \Lambda^2$, 
as clearly observed  experimentally~\footnote{See the review 69 on scalar mesons below 2~GeV in the Review of Particle Properties \cite{pdg}}.    

As a consequence, the dynamics of $S$-wave $\pi\pi$ scattering are strongly influenced by this resonance even at energies significantly lower than the cutoff of the EFT, and similar effects are expected \textit{mutatis mutandis} for the graviball in graviton-graviton scattering for $\omega\gtrsim 1$.


\section{Outlook}\label{sec:outlook}

Our paper presents the first evidence for a resonance in the sub-Planckian scattering of pure gravitons using well-established methods to unitarize the tree-level amplitudes. This result opens several interesting future directions worth pursuing.
Interesting future directions of investigation include unitarizing \emph{subleading} graviton-graviton scattering amplitudes. 
These processes at one and two-loop level are known in the literature with different matter content, e.g. \cite{Dunbar:1994bn,Abreu:2020lyk,Goroff:1985th,Bern:2020gjj}. This direction will allow us to complement the study we initiated for $R^4$ monomials in this paper, study the impact of matter fields and also  consider multiparticle cuts and other higher order effects in the EFT. In particular, higher order calculations will  reduce the dependence of our results on the IR regulator $\log a$.
Similarly, string theory amplitudes may be used for this purpose, see e.g.~\cite{Schwarz:1982jn,Alonso:2019ptb} or  
other inputs from other candidates to UV complete theories of gravitation, e.g.\cite{weinberg:safe,Niedermaier:2006wt,Codello:2008vh,Falls:2014tra,Blas:2009qj,Steinwachs:2020jkj}. 
Let us notice that our  unitarization formalism could be extended to  coupled channels involving other (massive) particle species \cite{han.200204.1,donoghue.200204.1,calmet.200204.1}. 
It is also interesting to deepen the connection between the graviball and the bound states investigated in the scattering of gravitons~\cite{Dvali:2014ila}.

Possible phenomenological signals of this scalar resonance may appear in situations with energies close to the fundamental energy scale of the gravitational theory. We commented in the main text that the examples from the $\sigma$ hint towards processes where these signals may appear. In a minimal scenario case, one can imagine that in the inflationary era, the \emph{graviball} may affect the dynamics of the inflaton itself from gravitational rescattering, or impact the distribution of the primordial gravitational waves generated at this time (in particular regarding the net polarization of this background). Furthermore, as discussed above, certain fundamental theories reach 
 gravitational cut-offs of energies as low as TeV \cite{ArkaniHamed:1998rs,Dvali:2001gx}. In this case, the resonance becomes narrower and less massive than the fundamental cut-off \cite{graviballlong}.
 This will  impact the phenomenology of gravitational interactions in the early universe. Maybe more relevant is the fact that these resonances could also leave a trace on some searches in accelerators, following in spirit the works of gravitational theories with low-energy cut-off \cite{Emparan:2003xu,Dimopoulos:2001hw}. We hope to come back to these fascinating prospects in future works.
 
\section*{Acknowledgements}

The authors are grateful to A.~Dobado, J. Donoghue, G. Dvali, D. G. Figueroa, J.~Penedones, S. Patil and M.~Herrero-Valea for valuable discussions. This work has been supported in part by the  FEDER (EU) and MEC (Spain) Grants FPA2016-77313-P, PGC2018-102016-A-I00,  and the ``Ram\'on y Cajal'' program RYC-2016-20672. The MICINN AEI (Spain) Grant PID2019-106080GB-C22/AEI/10.13039/501100011033 is also acknowledged. DB is supported by a `Ayuda Beatriz Galindo Senior' from the Spanish `Ministerio de Universidades', grant BG20/00228. The research leading of to these results has received funding from the Spanish Ministry of Science and Innovation (PID2020-115845GB-I00/AEI/10.13039/501100011033).

\appendix

\bibliography{gravball.bib}
\bibliographystyle{apsrev4-1}

\end{document}